\documentclass[12pt, preprint]{aastex}

\shorttitle{LBTAO Images of HR 8799}
\shortauthors{Skemer et al.}

\begin{document}

\title{First Light LBT AO Images of HR 8799 bcde at 1.6 and 3.3$\micron$: New Discrepancies between Young Planets and Old Brown Dwarfs\footnote{The LBT is an international collaboration among institutions in the United States, Italy and Germany. LBT Corporation partners are: The University of Arizona on behalf of the Arizona university system; Istituto Nazionale di AstroÞsica, Italy; LBT Beteiligungsgesellschaft, Germany, representing the Max-Planck Society, the Astrophysical Institute Potsdam, and Heidelberg University; The Ohio State University, and The Research Corporation, on behalf of The University of Notre Dame, University of Minnesota and University of Virginia.}}

\author{Andrew J. Skemer$^{1}$, 
Philip M. Hinz$^{1}$, 
Simone Esposito$^{2}$,
Adam Burrows$^{3}$,
Jarron Leisenring$^{4}$
Michael Skrutskie$^{5}$
Silvano Desidera$^{6}$,
Dino Mesa$^{6}$,
Carmelo Arcidiacono$^{2,7}$,
Filippo Mannucci$^{2}$,
Timothy J. Rodigas$^{1}$,
Laird Close$^{1}$,
Don McCarthy$^{1}$,
Craig Kulesa$^{1}$,
Guido Agapito$^{2}$,
Daniel Apai$^{1}$,
Javier Argomedo$^{2}$,
Vanessa Bailey$^{1}$,
Konstantina Boutsia$^{8,9}$,
Runa Briguglio$^{2}$,
Guido Brusa$^{8}$,
Lorenzo Busoni$^{2}$,
Riccardo Claudi$^{6}$,
Joshua Eisner$^{1}$,
Luca Fini$^{2}$,
Katherine B. Follette$^{1}$,
Peter Garnavich$^{10}$,
Raffaele Gratton$^{6}$,
Juan Carlos Guerra$^{8}$,
John M. Hill$^{8}$,
William F. Hoffmann$^{1}$,
Terry Jones$^{11}$,
Megan Krejny$^{11}$,
Jared Males$^{1}$,
Elena Masciadri$^{2}$,
Michael R. Meyer$^{4}$,
Douglas L. Miller$^{8}$,
Katie Morzinski$^{1}$,
Matthew Nelson$^{5}$,
Enrico Pinna$^{2}$,
Alfio Puglisi$^{2}$,
Sascha P. Quanz$^{4}$,
Fernando Quiros-Pacheco$^{2}$,
Armando Riccardi$^{2}$,
Paolo Stefanini$^{2}$,
Vidhya Vaitheeswaran$^{1}$,
John C. Wilson$^{5}$,
Marco Xompero$^{2}$
}

\affil{$^{1}$Steward Observatory, Department of Astronomy, University of Arizona, 933 N. Cherry Ave, Tucson, AZ 85721}
\affil{$^{2}$Istituto Nazionale di Astrofisica, Osservatorio Astrofisico di Arcetri Largo E Fermi 5 50125 Firenze, Italy}
\affil{$^{3}$Department of Astronomy, Princeton University, 4 Ivy Lane, Princeton, NJ 08544}
\affil{$^{4}$Institute for Astronomy, ETH Zurich, Wolfgang-Pauli-Strasse 27, CH-8093 Zurich, Switzerland}
\affil{$^{5}$Department of Astronomy, University of Virginia, 530 McCormick Road, Charlottesville, VA 22904}
\affil{$^{6}$Istituto Nazionale di Astrofisica, Osservatorio Astronomico di Padova, Vicolo dell' Osservatorio 5, I-35122, Padova, Italy}
\affil{$^{7}$Istituto Nazionale di Astrofisica, Osservatorio Astronomico di Bologna, Via Ranzani 1, 40127, Bologna, Italy}
\affil{$^{8}$Large Binocular Telescope Observatory, University of Arizona, 933 N. Cherry Ave, Tucson, AZ 85721}
\affil{$^{9}$Istituto Nazionale di Astrofisica, Osservatorio Astronomico di Roma, Via Frascati 33 00040, Rome, Italy}
\affil{$^{10}$Department of Physics, University of Notre Dame, 225 Nieuwland Science Hall, Notre Dame, IN 46556}
\affil{$^{11}$School of Physics and Astronomy, University of Minnesota, 116 Church Street S.E., Minneapolis, MN 55455}

\begin{abstract}
As the only directly imaged multiple planet system, HR 8799 provides a unique opportunity to study the physical properties of several planets in parallel.  In this paper, we image all four of the HR 8799 planets at H-band and 3.3$\micron$ with the new LBT adaptive optics system, PISCES, and LBTI/LMIRCam.  Our images offer an unprecedented view of the system, allowing us to obtain H and 3.3$\micron$ photometry of the innermost planet (for the first time) and put strong upper-limits on the presence of a hypothetical fifth companion.  We find that all four planets are unexpectedly bright at 3.3$\micron$ compared to the equilibrium chemistry models used for field brown dwarfs, which predict that planets should be faint at 3.3$\micron$ due to CH$_{4}$ opacity.  We attempt to model the planets with thick-cloudy, non-equilibrium chemistry atmospheres, but find that removing CH$_{4}$ to fit the 3.3$\micron$ photometry increases the predicted L' (3.8$\micron$) flux enough that it is inconsistent with observations.  In an effort to fit the SED of the HR 8799 planets, we construct mixtures of cloudy atmospheres, which are intended to represent planets covered by clouds of varying opacity.  In this scenario, regions with low opacity look hot and bright, while regions with high opacity look faint, similar to the patchy cloud structures on Jupiter and L/T transition brown-dwarfs.  Our mixed cloud models reproduce all of the available data, but self-consistent models are still necessary to demonstrate their viability.  

\end{abstract}

\section{Introduction\label{Introduction}}
Efforts are underway to characterize the first generation of directly imaged extra-solar planets.  A principal focus has been the HR 8799 planetary system \citep{2008Sci...322.1348M,2010Natur.468.1080M}, which with four planets, is currently unique as a directly-imaged multiple planet system.  Studying these planets simultaneously is particularly powerful given their connected formation histories and appearances.

HR 8799 is a young, A5V star with a $\lambda$ Boo deficiency of heavy metals and 3 distinct circumstellar dust structures \citep{2008Sci...322.1348M,1969AJ.....74..375C,1999AJ....118.2993G,2009ApJ...705..314S}.  There is some disagreement about the age of the system.  Traditional age-dating methods, such as galactic space motion and Hertzprung-Russell diagram position suggest that HR 8799 has an age of 20-160 Myr \citep{2006ApJ...644..525M,2008Sci...322.1348M,2010ApJ...716..417H,2011ApJ...732...61Z}, while astroseismology estimates are more consistent with $\sim$1 Gyr, which would make the planets significantly more massive brown dwarfs \citep{2010MNRAS.405L..81M}.  Interestingly, the dynamical stability of the planets themselves places upper-limits on the masses of the planets \citep{2010ApJ...710.1408F,2010ApJ...721L.199M,2012arXiv1201.0561S}, which directly converts to a young system age, based on the planets' photometry and evolutionary models \citep{1997ApJ...491..856B,2000ApJ...542..464C}.

An important implication of the relative youth and low-masses of the HR 8799 planets is that their appearances and atmospheric properties might be different than field brown dwarfs, which can have the same effective temperatures as giant planets while being older and more massive.  Brown dwarf spectra have been used as proxies for giant planet spectra to plan direct imaging surveys and to interpret early discoveries.  However initial results show that there are several key differences between the atmospheres of giant exoplanets and brown dwarfs.

For field brown-dwarfs, the L$\rightarrow$T spectral type transition, occurs at $\sim$1200-1400 K, where dust clouds settle/condense below the photosphere \citep{2008ApJ...689.1327S}, and CO is converted to CH$_{4}$ \citep{2003ApJ...596..587B,2002ApJ...564..466G}.  For the HR 8799 planets, clouds are suspended in the photosphere at lower effective temperatures (900-1200 K) than is typical for brown dwarfs \citep{2011ApJ...729..128C,2011ApJ...737...34M,2011ApJ...733...65B}.  Additionally, there appears to be more CO than CH$_{4}$ relative to equilibrium chemistry models, implying that convection is mixing hot material into the photosphere faster than the CO$\leftrightarrow$CH$_{4}$ reaction can re-equilibriate \citep{2010ApJ...716..417H,2011ApJ...733...65B}.  Similar, but more extreme results have been found for 2MASS 1207 b, a 5-7 $M_{jup}$ companion to a 25 $M_{jup}$ TW Hya brown-dwarf primary \citep{2004A&A...425L..29C,2011ApJ...732..107S,2011ApJ...735L..39B}.  Evidently, the HR 8799 planets and 2MASS 1207 b look similar to L-type brown-dwarfs, despite having effective temperatures more consistent with T-type brown-dwarfs.  

Multi-wavelength photometry and spectroscopy are the keys to understanding the differences between brown dwarfs and giant planets.  In particular, working over a broad wavelength range is critical for understanding clouds, chemistry, and the radiative budget of extrasolar planets.  The challenge of working over a broad wavelength range is that adaptive optics (AO) systems perform better at longer wavelengths where atmospheric turbulence is less severe.  But background radiation increases at longer wavelengths, and most AO systems have numerous warm optics, which can make working at long wavelengths impractical \citep{2000PASP..112..264L}.  The sweet-spot for most AO systems has typically been the near-infrared ($\sim$1-2.5$\micron$), although for extrasolar planets, there is a benefit to working at L' (3.8$\micron$) where the planet-star contrast improves \citep{2008ApJ...688..583H}.  

The Large Binocular Telescope (LBT) AO system can increase the wavelength range over which we study extrasolar planets.  With its 672-actuator deformable secondary mirror (installed on one side of the telescope at the time of our observations) the system produces unprecedented image quality and contrast at short wavelengths.  And because it is a deformable secondary AO system, it has a minimal number of warm optics, so that background noise remains low at long wavelengths.  

In this paper, we present Large Binocular Telescope (LBT) First Light Adaptive Optics (FLAO) images of HR 8799 at H-band (1.65$\micron$) and 3.3$\micron$, detecting the four planets (b-e) at both wavelengths.  Both images are superior to previous attempts at these wavelengths due to the high performance of the LBT's AO system.  Our images are the first detections of HR 8799 e at H-band and 3.3$\micron$ and the first unambiguous detections of HR 8799 b and d at 3.3$\micron$.  In Section 2, we give a basic description of the instrumental setup for the FLAO system, the near-infrared imager, PISCES, and the mid-infrared imager, LMIRCam, which is a component of LBTI.  Additionally, we describe our data reduction procedure, which is an implementation of the Locally Optimized Combination of Images algorithm (LOCI).  In Section 3 we estimate photometry for the four planets, based on our LOCI reductions.  In Section 4 we use our H-band image to search for additional companions interior to HR 8799 e, taking advantage of the unprecedented contrast afforded by the LBT AO system.  In Section 5 we present thick-cloud, non-equilibrium chemistry model atmospheres and mixed cloud atmospheres in an effort to explain the appearances of the HR 8799 planets.  We conclude in Section 6 and make suggestions for future work characterizing HR 8799 and other directly-imaged exoplanets.  A companion paper, \citet{2012arXiv1203.2735E}, describes the instrumental setup for the AO system and PISCES in detail, provides an independent analysis of the H-band data (along with new Ks-band data), and presents astrometry and a new orbital analysis of the system.

\section{Observations\label{Observations}}
\subsection{PISCES H-band\label{PISCES Obs}}
We observed HR 8799 at H-band ($\lambda=$1.66$\micron$; FWHM=0.29$\micron$) with PISCES \citep{2001PASP..113..353M} on UT Oct 16, 2011, during Science Verification Time for the LBT First-Light Adaptive Optics System (FLAO).  The LBT's FLAO system (PI-Simone Esposito) is a 672-actuator deformable secondary adaptive optics system that makes use of an innovative pyramid wavefront sensor, producing high Strehl-ratio, low background images over a broad wavelength range \citep{2010ApOpt..49G.174E,2011SPIE.8149E...1E}.  At the time of our observations, one adaptive optics system was installed on the right telescope, so for the observations presented in this paper, only one 8.4 meter mirror is used.  PISCES (PI-Don McCarthy) is a 1-2.5$\micron$ imager with a Hawaii 1024$\times$1024 HgCdTe array, which at the LBT (single 8.25 meter aperture), critically samples a diffraction-limited point-spread-function (PSF) at H-band (with a plate scale of 0.0193" per pixel).  For the observations in this paper, PISCES was installed on the front bent-Gregorian focus of the LBT's right side to take advantage of the LBT's new AO system while facility infrared cameras are being delivered and commissioned.

During our H-band observations of HR 8799, conditions were photometric and the natural seeing, as measured by a DIMM (Differential Image Motion Monitor) on the telescope structure, was as good as $\sim$0.9".  We obtained 901 images with 2-second integrations, over the course of 2 hours (90 degrees of sky-rotation) with the telescope rotator turned off \citep[ADI-Angular Differential Imaging;][]{2006ApJ...641..556M}.  PISCES' readout-time is 6 seconds, which means our observations were inefficient.  However, high-contrast observations are usually limited by PSF stability rather than photon-noise, so the long readouts had a negligible effect on our final results.  The 2-second integrations saturated out to a radius of $\sim$0.15".  We were not able to obtain unsaturated H-band images of the star for photometry and astrometry with PISCES' shortest integration time, 0.8 seconds.

Images were processed to remove cross-talk and persistence as described in \citet{2001PASP..113..353M} using the \textit{corquad} correction software\footnote{http://aries.as.arizona.edu/$\sim$observer/dot.corquad.pisces}.  We then dark-subtracted, flat-fielded, and distortion-corrected the images.  Finally, images were aligned by maximizing their cross-correlation.  We processed the aligned images using the LOCI algorithm \citep[Locally Optimized Combination of Images;][]{2007ApJ...660..770L}, which has been shown to produce higher contrast images than other algorithms, such as ADI.  In the terminology of \citet{2007ApJ...660..770L}, we used $N_{\delta}=1$, $N_{A}=300$, $g=1$, and a 1-pixel subtraction-region.\footnote{The general idea of LOCI with ADI is as follows: The stellar PSF is removed in a set of subtraction regions, which together, comprise the image.  For each subtraction region, the stellar PSF is estimated by constructing a linear combination of individual exposures that minimizes noise within a corresponding optimization region, which is centered on the subtraction region, but is much larger.  The optimization region is an annulus with an area of $N_{A}$ PSF-cores and a ratio between its radial extent and azimuthal extent of $g$. Images whose parallactic angle differ by less than an amount so that a source would move by $N_{\delta}$ FWHM are excluded from the optimization.}  The values for $N_{A}$ and $g$ are adopted from \citet{2007ApJ...660..770L} and we use $N_{\delta}=1$ instead of $N_{\delta}=0.5$ to suppress self-subtraction, although we note that changing $N_{\delta}$ to 0.5 has a negligible effect on our photometry.  Our implementation of the LOCI algorithm also includes a FWHM-sized mask around every subtraction-region, to further suppress self-subtraction.

\begin{figure}
\begin{center}
\includegraphics[angle=0,width=\columnwidth]{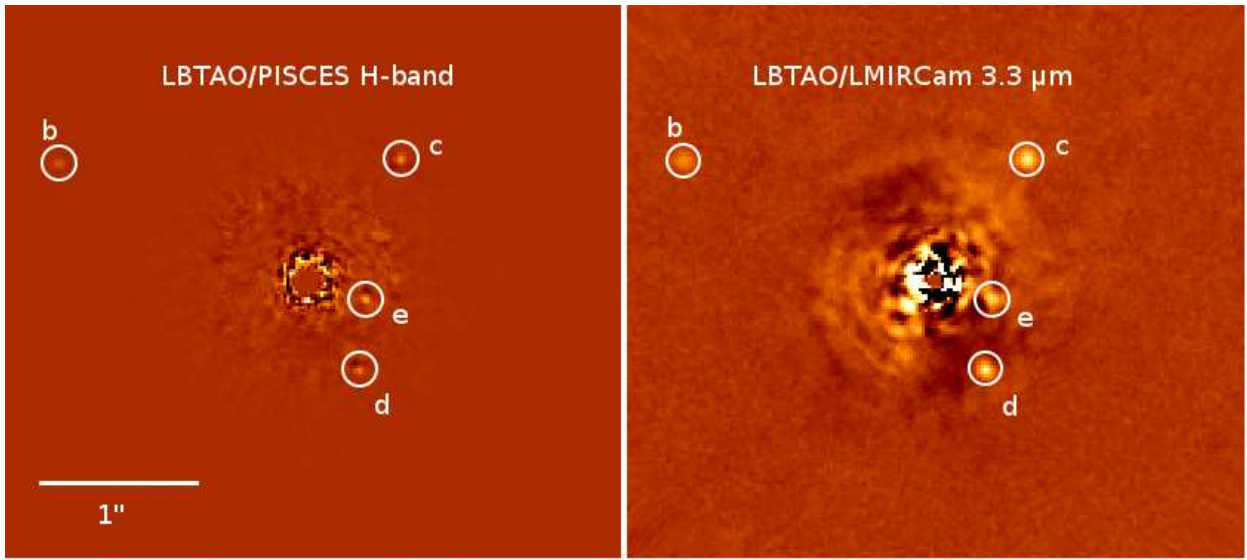}
\caption{LBT First Light AO images of the HR 8799 planetary system at H-band and 3.3$\micron$.  These images comprise the first detection of HR 8799 e at either wavelength, and the first unambiguous detections of HR 8799 b and d at 3.3$\micron$.
\label{Image}}
\end{center}
\end{figure}

After reducing all 901 images with LOCI, we evaluated the noise of the images (standard deviation of pixel count) within an annulus from 0.3"-0.5".  The noise dramatically increased after the ~500th image, corresponding to an increase in natural seeing.  We reran LOCI using just the first 500 images (1000 seconds, 61 degrees sky-rotation), marginally improving our final image, which is shown in Figure \ref{Image}.  HR 8799 b, c, d and e are all clearly visible.

The LOCI algorithm assumes accurate knowledge of the star's position.  However, the stellar core saturated even with PISCES' fastest readout.  We were able to estimate the stellar centroid to within 0.5 pixels (0.01") based on the circular symmetry of the PSF.  We then reran LOCI with a set of different stellar centroid positions comprising a 1$\times$1 pixel box and found no significant astrometric or photometric discrepancies between the results.

\subsection{LBTI/LMIRCam 3.3$\micron$\label{LMIRCAM Obs}}
We observed HR 8799 at 3.3$\micron$ ($\lambda=$3.31$\micron$; FWHM=0.40$\micron$) with LBTI/LMIRCam \citep[Large Binocular Telescope Interferometer/L and M-band Infrared Camera;][]{2008SPIE.7013E..67H,2010SPIE.7735E.118S} on UT Nov 16, 2011.  Although LBTI will eventually be used to combine the light from both LBT apertures, it was used in single-aperture mode for these observations, since only one adaptive optics system was operational.  LBTI (PI-Phil Hinz) consists of a beam combiner (UBC-Universal Beam Combiner), which combines the light from the two telescope tertiary mirrors, and a science camera (NIC-Nulling Infrared Camera), which itself contains a 2-5$\micron$ channel (LMIRCam) and an 8-13$\micron$ channel (NOMIC-Nulling Optimized Mid-Infrared Camera).  LMIRCam uses a Hawaii 2RG 2048$\times$2048 HgCdTe array, which oversamples $\lambda=$2-5$\micron$ PSFs (with a plate scale of 0.0106" per pixel) when using only one telescope primary.

Conditions during our 3.3$\micron$ observations were photometric, with a natural seeing (as measured by the telescope DIMM) of $\sim$1.1".  We obtained 1920 images with 3.8 second integrations over the course of 3.3 hours (110 degrees of sky rotation).  The 3.8 second exposure saturated inside of 0.06".  We also obtained short (0.15 second) unsaturated images for astrometry/photometry. About 1\% of the images had bad tip/tilt residuals and were removed.

Images were flat-fielded and globally-bad pixels (defined as pixels further than 15\% from the median flat) were replaced with the average of the 8-nearest good pixels.  The images were then nod-subtracted with nods taken every 20 images.  After these basic reduction steps, we found that some pixels had changing biases over shorter timescales than the nod-subtraction.  We removed these with three steps: subtracting the 3$\sigma$-clipped median of each column (to remove column bias-level effects), median-binning the data with a 2$\times$2 box (since the PSF is over-sampled by a factor of 3 for a single aperture telescope), and replacing isolated bad-pixels (4-sigma from their neighboring pixels) with the median of the neighboring pixels  For the remainder of this paper, LMIRCam ``pixels" refer to binned pixels, which have a plate-scale of 0.0212".

Images were processed with LOCI, using the same parameters and implementation described in Section \ref{PISCES Obs}.  We evaluated the noise inside a 0.2-0.4" annulus in each LOCI-processed image, and removed $\sim$10\% of the images, which had high noise, due to residual bad-pixels and/or sub-optimum AO performance.  Our final image, which is shown in Figure \ref{Image}, clearly shows planets b, c, d and e.

\section{Photometry\label{Photometry}}
Since LOCI self-subtracts, we calibrated our photometry by subtracting fake planets from the raw data at the positions of the detected planets, and rerunning the full LOCI pipeline.  Best-fit photometry and error bars were determined by adjusting the fluxes of the fake planets to determine the range of values resulting in reasonable subtractions.  For the PISCES H-band data, the star was saturated, so we calculated photometry for planets c, d and e with respect to HR 8799 b, and converted to absolute magnitudes by adopting the magnitude of HR 8799 b from \citet{2009ApJ...705L.204M}.  For the LMIRCam 3.3$\micron$ data, we calculated photometry for all four planets with respect to unsaturated images of the star, obtained immediately after the saturated images used to detect the planets.  We converted these to absolute photometry using HR 8799's absolute magnitude at 3.3$\micron$ from \citet{2010ApJ...716..417H}.  Errors on the LMIRCam absolute calibration are primarily the result of changing Strehl-ratios and telluric absorption variation throughout our observations.  To test the magnitude of the Strehl-ratio variation, we did $r=\lambda/D$ aperture photometry on the unsaturated data and found a standard deviation of only $\sim$2\%.  To test the magnitude of the telluric absorption variation, we compared the Airy pattern of the saturated and unsaturated images and found them to be consistent within $\sim$5\%.  Combining these error terms, we adopt an absolute calibration error of 0.06 mags for the LMIRCam 3.3$\micron$ data.  H-band and 3.3$\micron$ photometry of HR 8799 are presented in Table \ref{Photometry Table}.  

\begin{figure}
\begin{center}
\includegraphics[angle=0,width=\columnwidth]{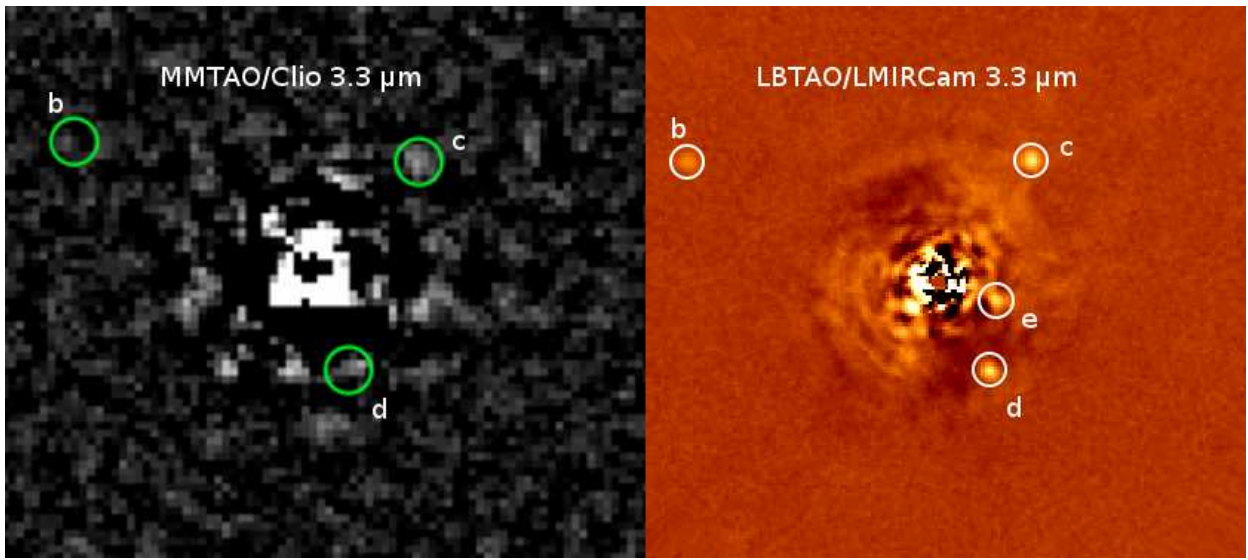}
\caption{A comparison of the MMTAO/Clio 3.3$\micron$ image from \citet{2010ApJ...716..417H} and our new LBTAO/LMIRCam 3.3$\micron$ image.  Our 3.3$\micron$ photometry is somewhat inconsistent with the findings of \citet{2010ApJ...716..417H} and \citet{2011ApJ...729..128C}, who separately analyzed the MMT data.  Based on the relative quality of the images, it is likely that the disparity is the result of overly-optimistic error bars by \citet{2010ApJ...716..417H} and \citet{2011ApJ...729..128C}.
\label{MMT_compare}}
\end{center}
\end{figure}

Our H-band photometry for HR 8799 c and d are consistent with the results of \citet{2008Sci...322.1348M} and \citet{2009ApJ...705L.204M}, and we detect HR 8799 e for the first time at H-band, finding it to be $\sim$0.3 mags brighter than the next brightest planet (HR 8799 c).  Our 3.3$\micron$ data are somewhat inconsistent with previous photometry from \citet{2010ApJ...716..417H} and \citet{2011ApJ...729..128C}, which are independent reductions of observations taken with MMT/Clio (note that the Clio and LMIRCam 3.3$\micron$ filters are identical).  \citet{2010ApJ...716..417H} reported detections of HR 8799 c and d but not b, and \citet{2011ApJ...729..128C} reported detections of HR 8799 b and c but not d.  The most substantial disparity is for HR 8799 b, for which \citet{2010ApJ...716..417H} reported an absolute magnitude upper-limit of 14.82, while \citet{2011ApJ...729..128C} reported a detection of 13.96$\pm$0.28.  Our detection of 13.22$\pm$0.11 is closer to the result of \citet{2011ApJ...729..128C} but is still brighter by a significant amount.  We reanalyzed the final reduced images from \citet{2010ApJ...716..417H} and find the upper-limit reported by \citet{2010ApJ...716..417H} to be erroneous (likely a typographic error).  Our photometry for HR 8799 c is also brighter than observed by \citet{2010ApJ...716..417H} and \citet{2011ApJ...729..128C} and our photometry of HR 8799 d is brighter than observed by \citet{2010ApJ...716..417H}.  We present a comparison of the MMT/Clio image and our new LBT/LMIRCam image in Figure \ref{MMT_compare}.  The fact that our photometry is somewhat inconsistent with the values of \citet{2010ApJ...716..417H} and \citet{2011ApJ...729..128C} can likely be explained by the non-photometric conditions reported in \citet{2010ApJ...716..417H}, varying AO performance (caused by the non-photometric conditions) and overly-optimistic systematic and/or measurement error analyses considering the quality of the Clio data.  Variability is unlikely to be a factor, given that it has not been reported in any other photometric band and the amplitude of variability needed to rectify the disparity is quite large.

\section{Constraints on a Hypothetical Fifth Planet, HR 8799 f}
While HR 8799 is known to have four giant planets at wide separations, the inner system might have one or more companions that have not yet been discovered.  Additional companions would challenge formation models, which already have a hard time explaining the outer four planets \citep{2009ApJ...707...79D,2010ApJ...710.1375K}.  A fifth planet would also complicate dynamical stability analyses, which require mean-motion resonances and planet masses that are on the low end of the range predicted by evolutionary models and independent age estimates \citep{2010ApJ...710.1408F,2010ApJ...721L.199M,2012arXiv1201.0561S}.  \citet{2011ApJ...730L..21H} used non-redundant masking to rule out the presence of massive inner companions from $\sim$0.01"-0.5", but only for objects significantly more massive than the four known planets.  Here we use our H-band image to search for planetary mass companions interior to HR 8799 e.

\begin{figure}
\begin{center}
\includegraphics[angle=0,width=\columnwidth]{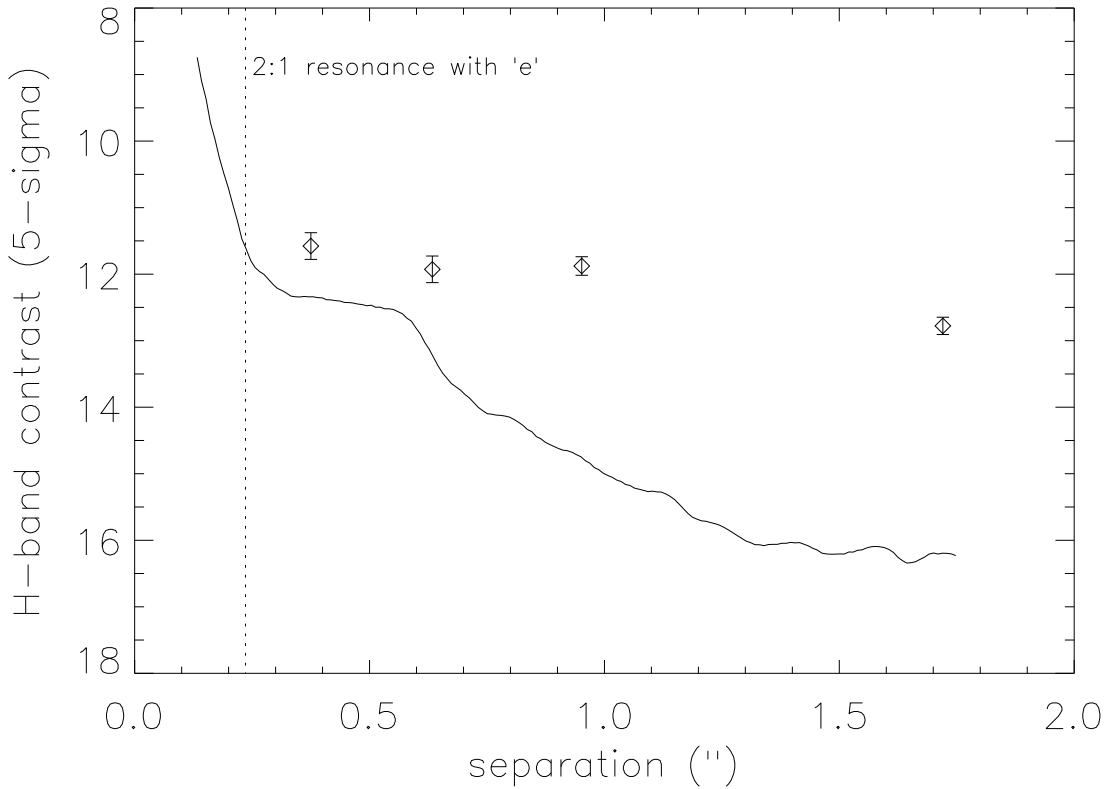}
\caption{H-band contrast curve for the LBTAO/PISCES H-band image of HR 8799, with the four planets shown.  Also shown is the position of the 2:1 orbital resonance with HR 8799 e (assuming, for simplicity, a face-on, non-eccentric orbit).  If there is a massive inner planet, it is likely to be in a stable resonance, as has been found for the outer companions.  Based on the contrast curve, we would have been able to detect a planet at HR 8799 e's 2:1 inner resonance, if it were approximately as bright (massive) as HR 8799 cde.  Note that the contrast curve shows a dark hole inside of $\sim$0.6", which is a predicted feature of high-order adaptive optics systems \citep{1995PASP..107..386M}.
\label{contrast}}
\end{center}
\end{figure}

\begin{figure}
\begin{center}
\includegraphics[angle=0,width=\columnwidth]{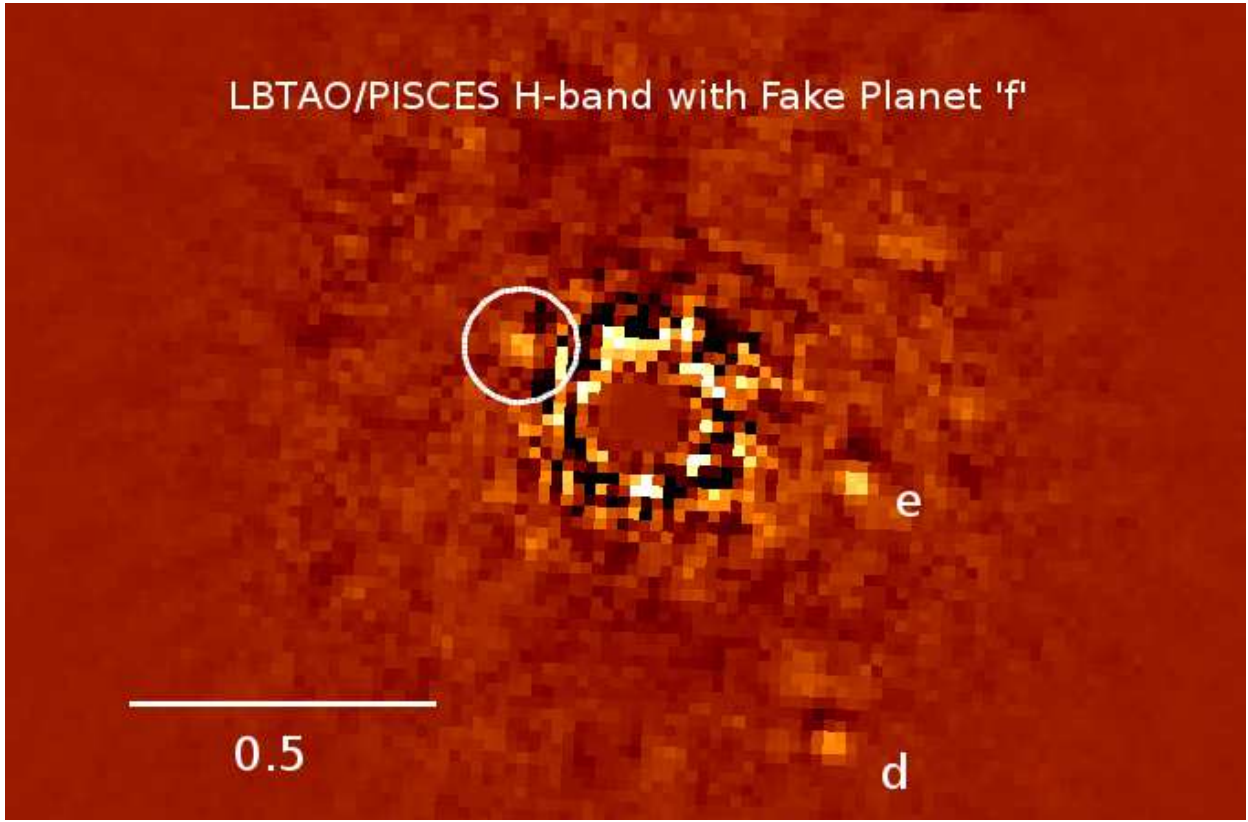}
\caption{Image of the HR 8799 system with a fake planet, `f', added in at the approximate location of HR 8799 e's 2:1 orbital resonance.  The fake planet, which is the same brightness as HR 8799 e, was added into our individual raw frames and is easily recovered by our LOCI pipeline.  There are no residual point-sources as bright as the fake planet at or exterior to its position at the HR 8799 e 2:1 orbital resonance.
\label{planet f}}
\end{center}
\end{figure}

We evaluate our ability to detect a close-in companion by making a contrast curve of our residual H-band image (after the 4 planets have been removed).\footnote{Note that we use an H-band image constructed with $N_{\delta}=0.5$ instead of $N_{\delta}=1.0$ (which was described in Section \ref{Photometry}) to maximize S/N, as described in \citet{2007ApJ...660..770L}.  The $N_{\delta}=0.5$ contrast curve is $\sim$0.1-0.2 mags better than the $N_{\delta}=1.0$ contrast curve from $\sim$0.2-0.3", and then becomes progressively worse (up to $\sim$0.5 mags) outside of 0.3".  For this section, the inner regions are more important, so we use the $N_{\delta}=0.5$ contrast curve.} We smooth the image with a Gaussian that is the same size as our diffraction-limited PSF, and calculate the standard deviation in 1-pixel annuli.  Counts are converted to photometry using the peak-flux (central pixel) of planet 'b' from the smoothed image.  We also correct for self-subtraction, which is measured by inserting fake planets into the raw data at various radii.  Our 5$\sigma$ contrast curve is shown in Figure \ref{contrast}.  HR 8799 b-e are shown as diamonds.  The vertical dashed line denotes the separation of the 2:1 mean-motion resonance with HR 8799 e (assuming a face-on circular orbit for simplicity).  If there is a massive interior planet, it is likely to be in a stable resonance, as has been found for pairs of the outer four planets.  We find no fifth planet at or exterior to the 2:1 resonance with HR 8799 e, with limits down to the approximate brightnesses (and by extension, masses) of the inner three planets.  As a check, we insert a fake `e'-like planet into our raw-images at a separation of 0.235", the approximate position of the 2:1 resonance with `e'.  Our reduced image (Figure \ref{planet f}) shows that we would have detected ``planet f" anywhere exterior to the 2:1 resonance, and that there are no residuals brighter than `f' in the image.  Note that we do not repeat this analysis with the 3.3$\micron$ image because it is not as sensitive to additional companions as the H-band image.

\section{Multiwavelength Modeling of the HR 8799 planets}
Multiwavelength photometry and spectroscopy have improved our understanding of the physical properties of the HR 8799 planets, which are not well fit by the same models that have been used to interpret the properties of field brown dwarfs.  \citet{2008Sci...322.1348M} first noted that the HR 8799 planet SEDs were best fit by model atmospheres with T=1400-1700 K, while their luminosities were more consistent with T=800-1000 K.  The disparity was driven by the planets' faint/red appearance with respect to models, and a lack of methane absorption in narrow-band 1.59/1.68 CH$_{4}$ photometry.  Subsequent 3.88-4.10$\micron$ spectroscopy of HR 8799 c was inconsistent with both COND \citep{2003A&A...402..701B} and DUSTY \citep{2000ApJ...542..464C} atmospheric models, which the authors interpreted as evidence for non-equilibrium chemistry \citep{2010ApJ...710L..35J}. 3-color photometry in the L and M bands were also inconsistent with equilibrium chemistry atmospheric models, and in particular showed a relative lack of methane absorption at 3.3$\micron$ \citep{2010ApJ...716..417H}.  H and K spectroscopy of HR 8799 b also showed a methane deficiency \citep{2010ApJ...723..850B,2011ApJ...733...65B}.  \citet{2011ApJ...729..128C} and \citet{2011ApJ...737...34M} were able to parameterize thick cloud models to fit multiwavelength photometry for HR 8799 bcd.  However, the models were not able to reproduce the 3.3$\micron$ photometry or the subsequent spectroscopy of \citet{2011ApJ...733...65B}, who fit both the photometry and spectroscopy of HR 8799 b with models that incorporated clouds and non-equilibrium chemistry.

\begin{figure}
\begin{center}
\includegraphics[angle=90,width=\columnwidth]{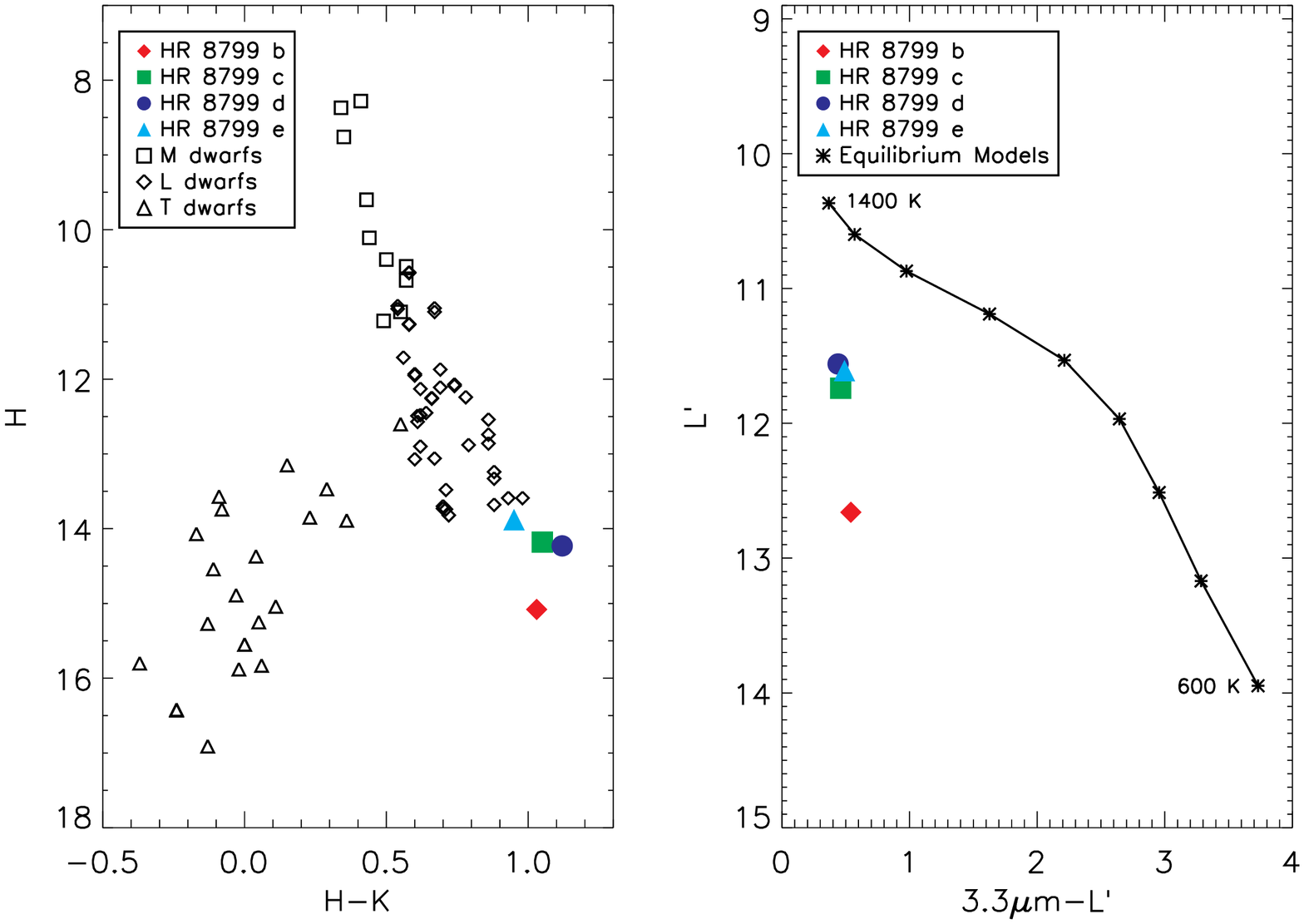}
\caption{LEFT: H vs. H-K color-magnitude diagram showing the M$\rightarrow$L$\rightarrow$T spectral-type transition for field brown dwarfs \citep{2002ApJ...564..452L,2004AJ....127.3553K} and the HR 8799 planets.  The HR 8799 planets appear to be an extension of the L-dwarf sequence, implying that they have cloudy atmospheres at lower effective temperatures than are typical for cloudy field brown dwarfs.
\newline
RIGHT: L' vs. 3.3$\micron$-L' color-magnitude diagram showing equilibrium chemistry, thick-cloud atmospheres from \citet{2011ApJ...737...34M} and the HR 8799 planets.  The HR 8799 planets are all brighter at 3.3$\micron$ than predicted by the \citet{2011ApJ...737...34M} models, implying a lack of CH$_{4}$, which is a strong absorber at 3.3$\micron$ in the equilibrium chemistry model atmospheres.  
\label{HK color mag}}
\end{center}
\end{figure}

The combined result of these studies is that HR 8799 b, c and d have non-equilibrium CO$\leftrightarrow$CH$_{4}$ chemistry and cloudy atmospheres at low effective temperatures where, in field brown dwarfs, the clouds are thought to have settled below the photosphere.  Two color-magnitude diagrams, shown in Figure \ref{HK color mag}, demonstrate these effects.  On the left, is an H-K versus H color-magnitude diagram, which shows the M$\rightarrow$L$\rightarrow$T sequence of field brown dwarfs.  L dwarfs are characterized by cloudy atmospheres, while T dwarfs are characterized by cloud-free atmospheres. The intermediate region is the L$\rightarrow$T transition, where objects are thought to have patchy clouds or clouds that have partially descended below the photosphere.  The HR 8799 planets appear to be an extension of the field L-dwarf sequence, i.e. they have clouds at an effective temperature where the field brown dwarfs are transitioning to cloudless.  The right-hand side of Figure \ref{HK color mag} shows a 3.3$\micron$-L' versus L' color magnitude diagram, which includes a sequence of chemical equilibrium, thick cloud models from \citet{2011ApJ...737...34M}.  The HR 8799 planets are all much brighter at 3.3$\micron$ then predicted by the models, implying a relative absence of CH$_{4}$, which is a strong absorber at 3.3$\micron$.  

In the following sections, we present model atmospheres of the HR 8799 planets in an attempt to reproduce their photometry. A summary of the available photometry is presented in Table \ref{all photometry}, colors are presented in Table \ref{colors}, and a listing of all the models used in this paper is presented in Table \ref{model names}.  For all model comparisons in this paper, we convolve the model planet atmosphere and a model of Vega \citep{1992AJ....104.1650C} with filter profiles to produce predicted magnitudes in the different filters, which are then compared to the measured photometry.  The filter profiles have all been multiplied by a model telluric atmosphere\footnote{http://www.gemini.edu/sciops/ObsProcess/obsConstraints/atm-models/cptrans\textunderscore zm\textunderscore 43\textunderscore 15.dat}.  For most filters, this step has a negligible effect, because the telluric atmosphere has a flat transmission profile, but Earth's atmosphere has a large transmission slope in the 3.3$\micron$ filter which changes the effective wavelength of the filter, and in turn, changes the model magnitudes by $\sim$0.1 mags.  For the sake of plotting the model fits, we convert the filter magnitudes to Jy by convolving the Vega model with filter profiles to produce zero-point fluxes.  For HR 8799 b, we also include the H and K spectroscopy from \citet{2011ApJ...733...65B}.  The HK spectrum comparison is made by smoothing the planet model atmosphere by the published spectrum's resolution (0.01$\micron$) and directly comparing to the observed spectrum.  The absolute fluxes of the H and K spectroscopy are tied to the H and K photometry, so for our comparison, we allow the overall brightness of the two spectra to vary, and only fit the shapes of the spectra.    

\subsection{HR 8799 b\label{b modeling}}
HR 8799 b is the most challenging atmosphere to explain because it is the coolest of the four planets and is thus the largest outlier in the color-magnitude diagrams shown in Figure \ref{HK color mag}.  Additionally, the HR 8799 b atmosphere is the most constrained of the four planets due to the H and K spectroscopy of \citet{2011ApJ...733...65B}.  Our new 3.3$\micron$ photometry of HR 8799 b is significantly brighter (by $\sim$100\%) than the previous published value by \citet{2011ApJ...729..128C}.  As described in Section \ref{Photometry}, the \citet{2011ApJ...729..128C} 3.3$\micron$ photometry is likely erroneous, due to non-photometric conditions and/or overly optimistic error bars.  There is the possibility that HR 8799 b is extremely variable at 3.3$\micron$, but we consider that scenario unlikely due to the many times HR 8799 has been observed at other wavelengths where there has been no evidence of such large variability.

Before proceeding with our modeling, we examine \citet{2011ApJ...733...65B}'s recent hypothesis regarding the SED of HR 8799 b.  \citet{2011ApJ...733...65B} were able to reproduce all existing photometry and spectroscopy of HR 8799 b by using a non-equilibrium CO$\leftrightarrow$CH$_{4}$ cloudy atmosphere with $T_{\rm eff}$=1100K and $log(g)$=3.5.  However, their model is inconsistent with interior evolutionary models, which predict a larger object radius.  \citet{2011ApJ...733...65B} also present model atmospheres that obey the predictions of the evolutionary tracks, with a best-fit model that has T=896 K and Z=1.0 metallicity.  This model demonstrates that it might be possible to explain HR 8799 b's appearance with a combination of clouds, non-equilibrium chemistry and higher than solar metallicity.  However, the model does not fit all of the existing photometry, and our new 3.3$\micron$ photometry makes the fit significantly worse.

\subsubsection{Non-equilibrium CO$\leftrightarrow$CH$_{4}$ Chemistry}
\begin{figure}
\begin{center}
\includegraphics[angle=90,width=\columnwidth]{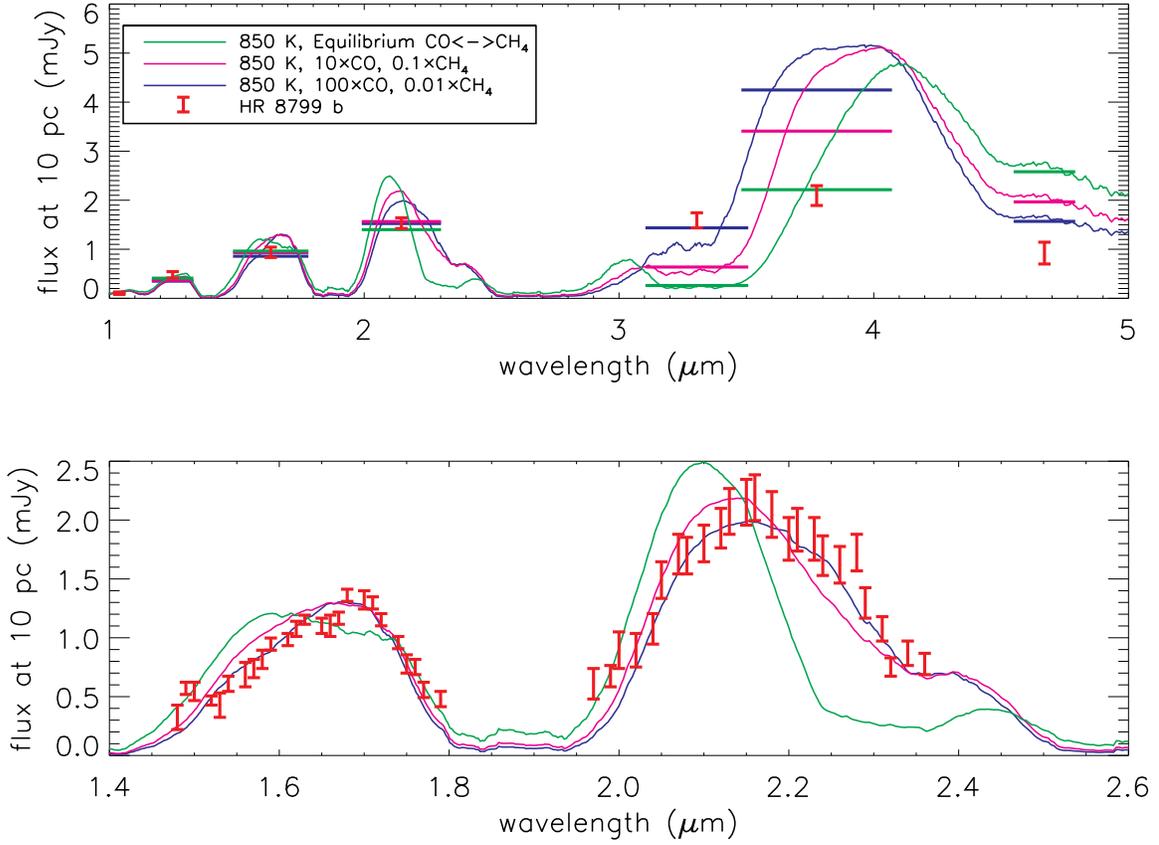}
\caption{Photometry and spectroscopy of HR 8799 b (red error bars; Table \ref{all photometry} and \citet{2011ApJ...733...65B}) with atmospheric models (green, pink and blue curves) and predicted in-band fluxes for each photometry point (green, pink and blue horizontal lines, which span the filters' half-max profiles).  The 850 K equilibrium chemistry model (green) is the best-fit ``thick-cloud" model for HR 8799 b from \citet{2011ApJ...737...34M}.  Two other models (pink and blue) suppress the CH$_{4}$ mixing ratios and enhance the CO mixing ratios by 10$\times$ and 100$\times$ with respect to the equilibrium chemistry model.  The H and K spectroscopy are well-fit by the non-equilibrium chemistry models.  Our new 3.3$\micron$ photometry is best fit by the 100$\times$CO, 0.01$\times$CH$_{4}$ model.  However, this model predicts a flux that is substantially higher than published measurements at 3.8$\micron$ (L').  None of the models are able to reproduce the relatively flat SED from 3.3-3.8$\micron$.
\label{non-eq chemistry}}
\end{center}
\end{figure}

We start our modeling by using the parameterized thick-cloud atmospheres of \citet{2011ApJ...737...34M} and adjusting the CO and CH$_{4}$ mixing ratios.  These models assume evolutionary track radii, but make no attempt to self-consistently explain the CO and CH$_{4}$ mixing ratios, which \citet{2011ApJ...733...65B} produce with turbulent mixing.  In Figure \ref{non-eq chemistry}, we plot the best fit chemical equilibrium model of HR 8799 b from \citet{2011ApJ...737...34M}, as well as two models that have suppressed CH$_{4}$ and enhanced CO with respect to the equilibrium models (by factors of 10 and 100 respectively).  Suppressing CH$_{4}$ and enhancing CO has a negligible effect on the near-infrared (zJHK) photometry, but it greatly affects the HK spectroscopy, favoring the 100$\times$CO, 0.01$\times$CH$_{4}$ model.  Non-equilibrium chemistry dramatically affects the 3-5$\micron$ SED, where a lack of CH$_{4}$ makes the object brighter in the 3.3$\micron$ and L' filters and excess CO makes the object fainter in the M-band filter.  The 100$\times$CO, 0.01$\times$CH$_{4}$ model fits the 3.3$\micron$ photometry, and comes closest to fitting the M-band photometry (further increasing the CO mixing ratio would improve this fit).  However, the lack of CH$_{4}$ strongly increases the predicted flux in the L' filter, making it incompatible with HR 8799 b's observed L' photometry.  Based on this analysis, it seems unlikely that a cool (850 K) atmosphere with non-equilibrium chemistry could explain HR 8799 b's 3.3$\micron$-L' color.  However, there remains the possibility that more complex radial profiles of non-equilibrium chemistry and clouds, could explain all of the data.

\subsubsection{Mixed Cloud Atmospheres\label{mixed b}}
Based on Figure \ref{HK color mag}, HR 8799 b has colors reflective of a $\sim$1300 K atmosphere, but a luminosity consistent with an $\sim$850 K atmosphere.  In lieu of resolving the difference by assuming a small (unphysical) radius, we consider the possibility that the planet emits non-isotropically, with bright and dark sections, such that the bright sections dominate the shape of the SED.  Bright and dark regions have been observed on Jupiter in the mid-infrared \citep{1969ApJ...157L..63W}, and several studies seeking to explain the L$\rightarrow$T transition have used hybrid cloudy/cloud-free models \citep[e.g.][]{2002ApJ...571L.151B}.  For our purposes, mixtures of standard cloudy and cloud-free models are unlikely to explain the appearance of HR 8799 b because its H-K color is redder than the L-dwarf sequence implying that it is more cloudy, not less cloudy, than the other L-dwarfs.  

\begin{figure}
\begin{center}
\includegraphics[angle=90,width=\columnwidth]{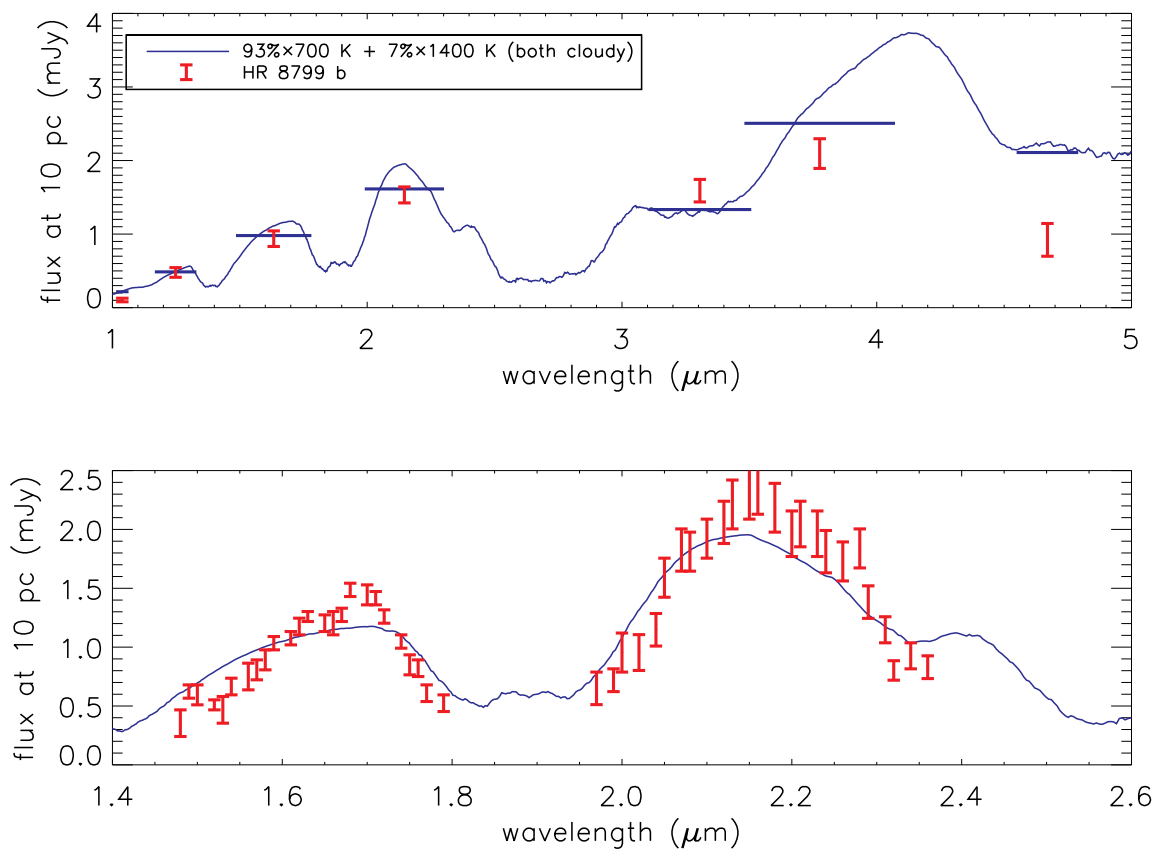}
\caption{Same as Figure \ref{non-eq chemistry} but with a mixed cloud model (93\% 700 K A-type clouds and 7\% 1400 K AE-type clouds from \citet{2011ApJ...737...34M}).  The fit adequately reproduces all photometry (except in the M-band filter, where additional CO absorption would rectify the discrepancy), and greatly flattens the SED from 3.3-3.8$\micron$ compared to the non-equilibrium models shown in Figure \ref{non-eq chemistry}.  At H and K, the model generally reproduces the shape of HR 8799 b's observed spectrum (i.e. negligible CH$_{4}$ absorption), but is flatter.
\label{hybrid cloud model}}
\end{center}
\end{figure}

An alternative is that HR 8799 b has a mixture of clouds, such that the whole planet is cloudy, but with regions that have thicker clouds where the planet appears darker.  A coarse approximation of this phenomenon is to linearly combine cloudy models of different effective temperatures (analogously to how \citet{2002ApJ...571L.151B} and others have linearly combined cloudy and cloud-free models of different effective temperatures).  This method is somewhat non-physical due to the fact that the two atmospheres have different temperature-pressure profiles \citep{2010ApJ...723L.117M}.  Linearly combining models with a shared temperature-pressure profile but different cloud structures would be more correct, but is beyond the scope of this paper.  In Figure \ref{hybrid cloud model}, we present an example of a hybrid atmosphere that is 93\% $T_{\rm eff}$=700 K, `A'-type cloudy and 7\% $T_{\rm eff}$=1400 K, `AE'-type cloudy (A and AE cloud profiles are described in \citet{2011ApJ...737...34M} and further model details are presented in \citet{2006ApJ...640.1063B}).  The hybrid model adequately fits all of the photometry except for M-band, which can be explained with further increased CO absorption.  The HK model spectrum is muted compared to the data, but its bulk shape is generally correct (i.e. it does not show strong CH$_{4}$ absorption, as would be expected for a cool object).  We note that this model is meant to be representative, but that a true mixed-cloud atmosphere should be calculated self-consistently.

\subsection{HR 8799 c, d and e}
\begin{figure}
\begin{center}
\includegraphics[angle=90,width=\columnwidth]{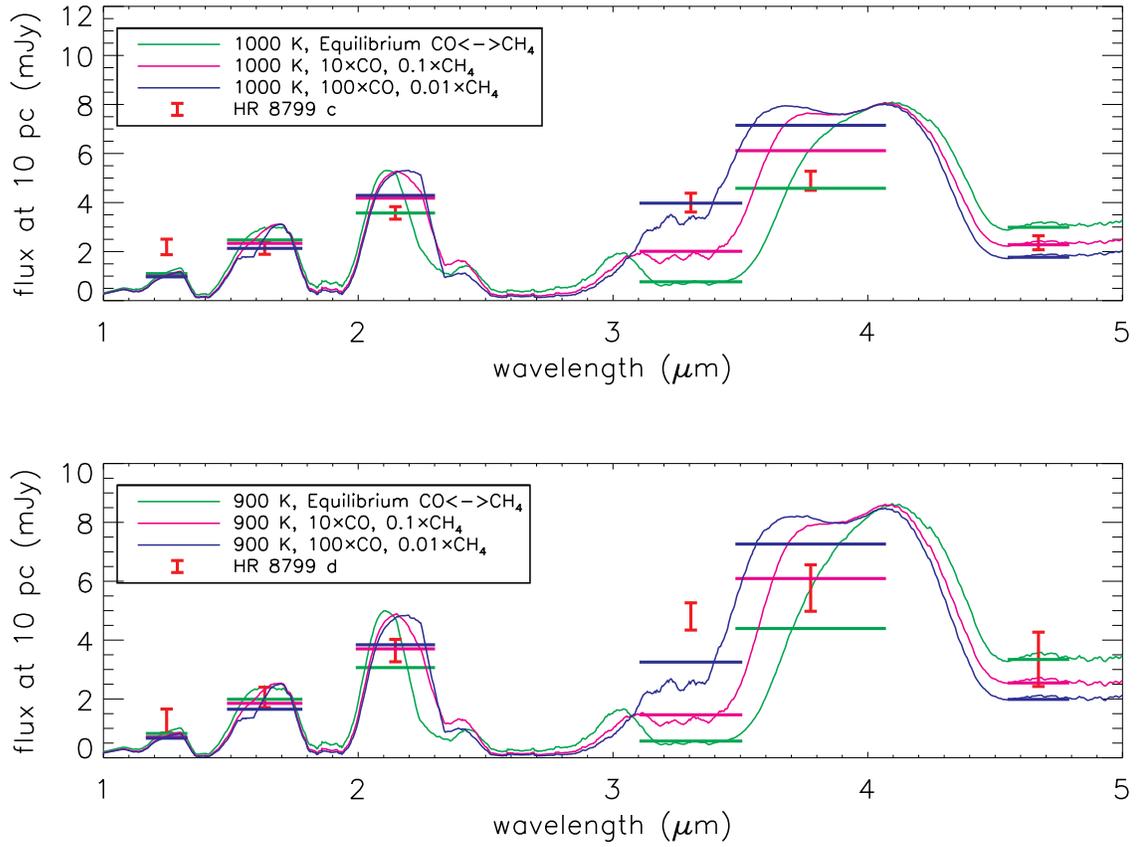}
\caption{Same as Figure \ref{non-eq chemistry} (top) but for HR 8799 c and d.  As was found for HR 8799 b, our non-equilbrium chemistry models are unable to fit the 3.3$\micron$-L' colors of HR 8799 c and d.
\label{cd non-eq}}
\end{center}
\end{figure}

\begin{figure}
\begin{center}
\includegraphics[angle=90,width=\columnwidth]{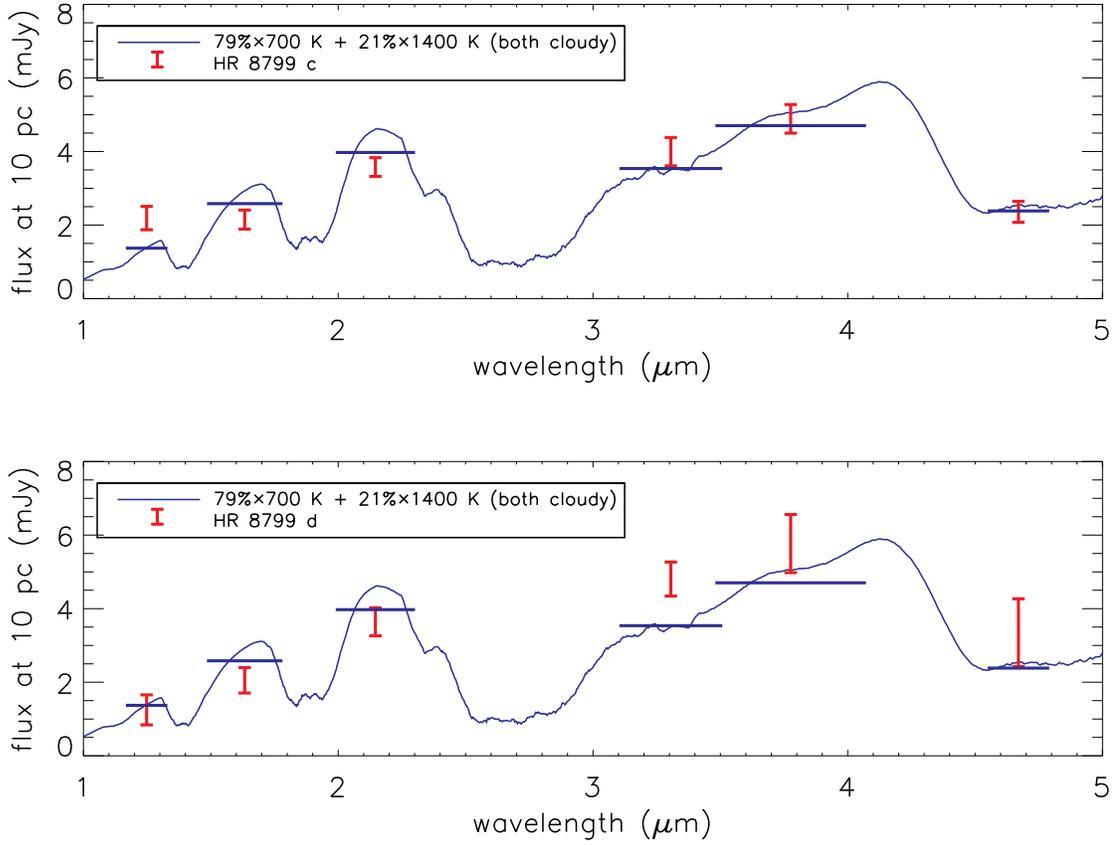}
\caption{Same as Figure \ref{hybrid cloud model} (top) but for HR 8799 c and d.  We purposely choose the same cloudy atmospheres to mix as were used for HR 8799 b.  We are able to fit HR 8799 c and d with mixed cloud atmospheres, but using a higher mixing fraction of the 1400 K atmosphere than was found for HR 8799 b.  The addition of non-equilibrium chemistry to these models would likely improve the fit.
\label{cd mixed}}
\end{center}
\end{figure}

We repeat our analysis of HR 8799 b (Section \ref{b modeling}) for HR 8799 c and d, again starting with the best-fit, thick-cloud models from \citet{2011ApJ...737...34M}.  A comparison of non-equilibrium CO$\leftrightarrow$CH$_{4}$ models is shown in Figure \ref{cd non-eq} and example mixed cloud atmospheres are shown in Figure \ref{cd mixed}.  Our conclusions for HR 8799 c and d are similar to our conclusions for HR 8799 b: we are unable to fit the 3.3$\micron$-L' colors of HR 8799 c and d with cloudy/non-equilibrium chemistry models, and mixed cloud atmospheres do a reasonable job fitting all of the data.  We purposely construct the mixed cloud atmospheres from the same two model atmospheres used to make the mixture for HR 8799 b (see Section \ref{mixed b}).  In this scenario, we find that HR 8799 c and d have higher fractions of the warm atmosphere than HR 8799 b, explaining their higher luminosity.

\begin{figure}
\begin{center}
\includegraphics[angle=90,width=\columnwidth]{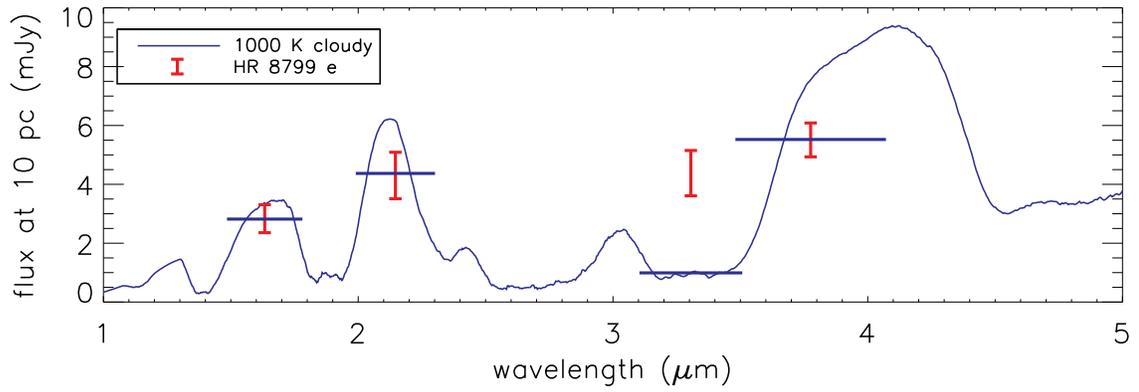}
\vspace{-185pt}
\caption{Thick cloud atmosphere for HR 8799 e (plotted with the same symbols used in Figures \ref{non-eq chemistry}-\ref{cd mixed}).  In this fit, we ignore the 3.3$\micron$ photometry as was done by \citet{2011ApJ...737...34M} when fitting HR 8799 b, c and d.
\label{e normal}}
\end{center}
\end{figure}

\begin{figure}
\begin{center}
\includegraphics[angle=90,width=\columnwidth]{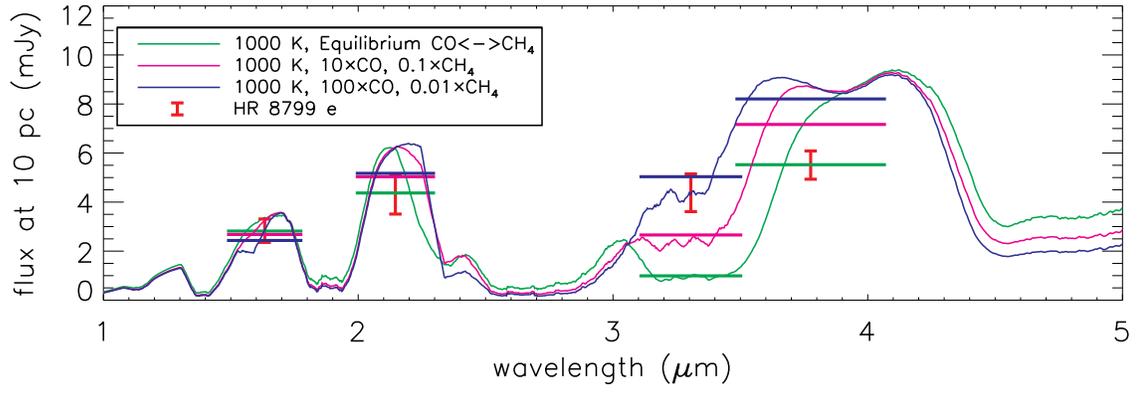}
\vspace{-185pt}
\caption{Same as Figures \ref{non-eq chemistry} (top) and \ref{cd non-eq} but for HR 8799 c and d.  As was found for HR 8799 b, c and d, our non-equilbrium chemistry models are unable to fit the 3.3$\micron$-L' colors of HR 8799 e.
\label{e non-eq}}
\end{center}
\end{figure}

\begin{figure}
\begin{center}
\includegraphics[angle=90,width=\columnwidth]{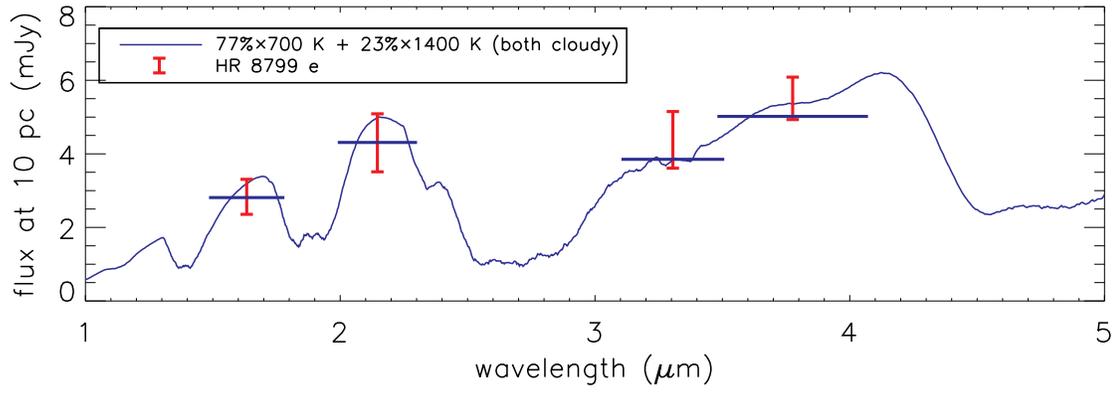}
\vspace{-185pt}
\caption{Same as Figures \ref{hybrid cloud model} (top) and \ref{cd mixed} but for HR 8799 e.  The mixing fraction of the two atmospheres is similar to what we found for HR 8799 c and d.  The addition of non-equilibrium chemistry to these models would slightly improve the fit, although the photometric error bars for HR 8799 e are large enough that this is not necessary.
\label{e mixed}}
\end{center}
\end{figure}

HR 8799 e has less data than the outer 3 planets, but the existing data is consistent with the photometry of HR 8799 c and d.  With the addition of our new H-band and 3.3$\micron$ photometry, HR 8799 e has now been studied at four wavelengths, and we can proceed with atmospheric modeling, based on lessons learned from HR 8799 b, c and d.  We begin by using the thick-cloud models from \citet{2011ApJ...737...34M} to look for a good fit of the H, K and L' photometry.  We find that a 1000 K, log(g)=4.0 model fits well (shown in Figure \ref{e normal}).  Based on the small number of data points, degeneracies between different cloud properties, surface gravity and effective temperature are not explored, but it is reasonable to assume the cloud properties and surface gravity will be similar to the other HR 8799 planets.  Figures \ref{e non-eq} and \ref{e mixed} show non-equilibrium chemistry models and mixed cloud models.  As was true for the outer three planets, we are unable to fit the available photometry with non-equilibrium chemistry models, and a mixed-cloud model can be made to fit.

\subsection{The HR 8799 Planets in Aggregate}

The HR 8799 system provides a unique laboratory for simultaneously studying multiple coeval planets.  In the previous sections, we have modeled the planets individually.  Comparing the four planets provides additional insight.  

HR 8799 c, d and e are brighter than HR 8799 b in all published photometry, but have almost exactly the same colors as `b' throughout their measured SEDs (see Table \ref{colors}; the colors for all four planets are the same, within errors, except for J-H for HR 8799 c, where J is abnormally bright).  Given that the planets are coeval, they are expected to all have similar radii \citep[to within $\sim$10\%;][]{1997ApJ...491..856B}.  Therefore, the fact that `c', `d' and `e' are brighter than `b' implies that they have higher effective temperatures (based on $L=4 \pi R^{2} \sigma T_{eff}^{4}$).  However, the fact that all of the HR 8799 planets have the same colors suggests that the physical properties of their atmospheres are similar, despite their different effective temperatures.  In field brown dwarfs, objects with different effective temperatures (over the range probed by the HR 8799 planets) have different physical properties and different SED colors.  For the HR 8799 planets, this does not appear to be the case.\footnote{As an example, \citet{2012arXiv1201.2465D} find $\Delta$K-L'/$\Delta$K=0.4 over a large effective temperature range, whereas HR 8799 b has a $\Delta$K-L'/$\Delta$K=0.03 with respect to the average values of HR 8799 c, d and e}

The fact that `c', `d' and `e' have almost the exact same colors as `b' is circumstantial evidence for the mixed-cloud atmospheres.  In the mixed-cloud scenario, HR 8799 b has a lower \textit{effective} temperature than the other HR 8799 planets, but a similar \textit{physical} temperature in the bright, emitting regions of its atmosphere.  In Figures \ref{hybrid cloud model}, \ref{cd mixed} and \ref{e mixed}, we have have shown that all of the HR 8799 planets can be fit by a mixture of 1400 K and 700 K cloudy atmospheres (which we have been using as an approximation for two different cloud structures, one of which has much lower opacity).  The difference between the cooler planet, `b', and the hotter planets, `c', `d' and `e', is the mixing ratio of the 1400 K and 700 K models, i.e. the covering fraction of the different cloud types.  

The non-equilibrium chemistry models (with reasonable radii) of \citet{2011ApJ...733...65B} fit the HK spectroscopy better than our makeshift mixed-cloud models, but they do not reproduce the broad-band colors as well, in particular from 3.3-3.8$\micron$.  The \citet{2011ApJ...733...65B} models do provide physical motivations for non-equilibrium chemistry (turbulent radial mixing), while our mixed cloud models are based on analogy with physically quite different systems (Jupiter and L/T transition brown-dwarfs).  Self-consistent modeling is still necessary to determine if mixed cloud atmospheres are a viable explanation for the HR 8799 planets.  In any case, it appears detailed modeling of increasingly complex cloud physics and chemistry will be necessary to explain the true nature of the HR 8799 planets.

\section{Summary and Conclusions}
We have directly imaged the HR 8799 planetary system, detecting all four planets at H-band and 3.3$\micron$ with the LBT's First Light Adaptive Optics system.  The images are of unprecedented quality allowing us to rule out the presence of a massive (HR 8799 cde-like) planet exterior to HR 8799 e's 2:1 inner resonance (H-band 5-$\sigma$ contrast of 11.6 magnitudes at 0.235").  We detect HR 8799 e at H-band, for the first time, and find that it is approximately as bright as HR 8799 c and d.  Combined with Ks and L' data \citep{2010Natur.468.1080M} and our new 3.3$\micron$ data, this indicates that HR 8799 e has similar atmospheric properties to HR 8799 c and d.  The planets are all brighter than expected at 3.3$\micron$, where equilibrium chemistry models predict CH$_{4}$ opacity should make the planets faint.

We model the HR 8799 planets with thick-cloudy atmospheres \citep{2011ApJ...737...34M} and allow the CO and CH$_{4}$ mixing ratios to vary arbitrarily.  The models that fit our 3.3$\micron$ data (which have very little CH$_{4}$) substantially over-predict the planets' fluxes at L' (3.8$\micron$).  Hotter atmospheres ($>$1300 K) have a similar 3.3$\micron$-3.8$\micron$ color as the HR 8799 planets, and the \textit{shape} of the \citet{2011ApJ...733...65B} HK spectrum is also well fit by an atmosphere that is significantly hotter than indicated by the HR 8799 planets' bolometric magnitudes.  As a result, we consider the possibility that small sections of the planets' atmospheres are hot ($>$1300 K), dominating the shape of the SEDs, while the majority of the planets' atmospheres are cooler and do not produce much flux.  The temperature-pressure profile must be the same between the ``hot" and ``cool" regions, so the physical difference would be that the ``cool" regions have increased cloud opacity.  Since the SED is consistent with a cloudy atmosphere, the ``hot" regions must also be cloudy, so the combined atmosphere is comprised of mixed clouds, some of which are thicker than others.  Our mixed cloud models are able to fit all of the HR 8799 data.  However, we caution that our models are not fully self-consistent and that more theoretical work is necessary to validate our hypothesis.  

The HR 8799 planets have unusual SEDs that are not well-fit by the same models that have been used to fit field brown dwarfs.  From an observational standpoint, HK spectroscopy and 3.3$\micron$-L' colors have been particularly powerful in ruling out model atmospheres.  HK spectroscopy of HR 8799 cde, and other directly-imaged planets in general, is critical to our understanding of clouds and non-equilibrium chemistry.  We also note that given the wide range of 3-4$\micron$ SEDs predicted by the models in this paper, it would be very useful to obtain low-resolution spectroscopy of the HR 8799 planets in this range.  From a theoretical standpoint, our new 3.3$\micron$ photometry is a challenge even for non-equilibrium chemistry models, which predict bright 3.3$\micron$ photometry.  Mixed cloud models are one possible way to flatten out 3.3$\micron$-3.8$\micron$ photometry and hide CH$_{4}$ opacity.

\acknowledgements
The authors thank Piero Salinari for his insight, leadership and persistence which made the development of the LBT adaptive secondaries possible.  We also thank the many dedicated individuals who have worked on LBTI, LMIRCam, PISCES, and the AO system over the years.  The Large Binocular Telescope Interferometer is funded by the National Aeronautics and Space Administration as part of its Exoplanet Exploration program.  LMIRCam is funded by the National Science Foundation through grant NSF AST-0705296.

\clearpage

\begin{deluxetable}{lcccccccccccc}
\tabletypesize{\scriptsize}
\tablecaption{LBT Photometry of the HR 8799 Planets}
\tablewidth{0pt}
\tablehead{
\colhead{Planet} &
\colhead{$\Delta$H Mag with HR 8799 b} &
\colhead{Absolute H Mag} &
\colhead{$\Delta$3.3$\micron$ Mag with HR 8799} &
\colhead{Absolute 3.3$\micron$ Mag}
}

\startdata
HR 8799 b &                & 15.08$\pm$0.13\tablenotemark{a} & 10.97$\pm$0.10 & 13.22$\pm$0.11 \\
HR 8799 c & -0.90$\pm$0.05 & 14.18$\pm$0.14                  & 9.97$\pm$0.10  & 12.22$\pm$0.11 \\
HR 8799 d & -0.85$\pm$0.2  & 14.23$\pm$0.2                   & 9.77$\pm$0.10  & 12.02$\pm$0.11 \\
HR 8799 e & -1.2$\pm$0.2   & 13.88$\pm$0.2                   & 9.87$\pm$0.20  & 12.02$\pm$0.21 \\

\enddata
\tablenotetext{a}{HR 8799 b absolute H-band photometry from \citet{2009ApJ...705L.204M}.  Absolute H-band photometry for the other planets is with respect to HR 8799 b.}
\label{Photometry Table}
\end{deluxetable}

\clearpage

\begin{deluxetable}{lcccccccccccc}
\rotate
\tabletypesize{\scriptsize}
\tablecaption{Photometry of the HR 8799 Planets}
\tablewidth{0pt}
\tablehead{
\colhead{Planet} &
\colhead{z} &
\colhead{J} &
\colhead{H} &
\colhead{$\rm{CH}_{4}\rm{s}$} &
\colhead{$\rm{CH}_{4}\rm{l}$} &
\colhead{Ks} &
\colhead{3.3} &
\colhead{L'} &
\colhead{M}
\\
\colhead{} &
\colhead{(1.03$\micron$)} &
\colhead{(1.25$\micron$)} &
\colhead{(1.63$\micron$)} &
\colhead{(1.59$\micron$)} &
\colhead{(1.68$\micron$)} &
\colhead{(2.15$\micron$)} &
\colhead{(3.3$\micron$)} &
\colhead{(3.8$\micron$)} &
\colhead{(4.7$\micron$)} &

}

\startdata
HR 8799 b                  & 18.24$\pm$0.29 & 16.30$\pm$0.16 & 15.08$\pm$0.13 & 15.18$\pm$0.17 & 14.89$\pm$0.18 & 14.05$\pm$0.08 & 13.2$\pm$0.11 & 12.66$\pm$0.11 & 13.07$\pm$0.30 \\
HR 8799 c                  &                & 14.65$\pm$0.17 & 14.18$\pm$0.14 & 14.25$\pm$0.19 & 13.90$\pm$0.19 & 13.13$\pm$0.08 & 12.2$\pm$0.11 & 11.74$\pm$0.09 & 12.05$\pm$0.14 \\
HR 8799 d                  &                & 15.26$\pm$0.43 & 14.23$\pm$0.2  & 14.03$\pm$0.30 & 14.57$\pm$0.23 & 13.11$\pm$0.12 & 12.0$\pm$0.11 & 11.56$\pm$0.16 & 11.67$\pm$0.35 \\
HR 8799 e                  &                &                & 13.88$\pm$0.2  &                &                & 12.93$\pm$0.22 & 12.1$\pm$0.21 & 11.61$\pm$0.12 &                \\
reference (b,c,d,e)        & 2              & 3,3,3          & 4,1,1,1        & 3,3,3          & 3,3,3          & 3,3,3,5        & 1,1,1,1 & 3,3,3,5        & 6,6,6          \\

\enddata

\tablerefs{
(1) This work;
(2) \citet{2011ApJ...729..128C};
(3) \citet{2008Sci...322.1348M};
(4) \citet{2009ApJ...705L.204M};
(5) \citet{2010Natur.468.1080M};
(6) \citet{2011ApJ...739L..41G}
}

\label{all photometry}
\end{deluxetable}

\clearpage

\begin{deluxetable}{lcccccccccccc}
\tabletypesize{\scriptsize}
\tablecaption{Colors of the HR 8799 Planets}
\tablewidth{0pt}
\tablehead{
\colhead{Planet} &
\colhead{J-H} &
\colhead{H-Ks} &
\colhead{Ks-L'} &
\colhead{3.3$\micron$-L'} &
\colhead{L'-M}
}

\startdata
HR 8799 b   & 1.22$\pm$0.20 & 1.03$\pm$0.15 & 1.39$\pm$0.13 & 0.54$\pm$0.15 & -0.41$\pm$0.31 \\
HR 8799 c   & 0.47$\pm$0.21 & 1.05$\pm$0.16 & 1.39$\pm$0.12 & 0.46$\pm$0.14 & -0.31$\pm$0.16 \\
HR 8799 d   & 1.03$\pm$0.46 & 1.12$\pm$0.23 & 1.55$\pm$0.20 & 0.44$\pm$0.19 & -0.11$\pm$0.37 \\
HR 8799 e   &               & 0.95$\pm$0.29 & 1.32$\pm$0.25 & 0.49$\pm$0.24 &                \\
\enddata
\label{colors}
\end{deluxetable}

\clearpage

\begin{deluxetable}{ccccccccccccc}
\tabletypesize{\scriptsize}
\tablecaption{Atmospheric Models Used in This Paper}
\tablewidth{0pt}
\tablehead{
\colhead{Figure \#} &
\colhead{T$_{eff}$ (K)} &
\colhead{log(g)} &
\colhead{Cloud Type} &
\colhead{Chemistry}
}

\startdata
\ref{non-eq chemistry}    & 850  & 4.3 & AE.60 & equilibrium                          \\
                          &      &     &       & 10 $\times$CO, 0.1$\times$CH$_{4}$   \\
                          &      &     &       & 100$\times$CO, 0.01$\times$CH$_{4}$  \\
\hline
\ref{hybrid cloud model}  & 1400 & 4.0 & AE.60 & equilibrium                          \\
                          & 700  &     & A.100 &                                      \\
\hline
\ref{cd non-eq}           & 1000 & 4.2 & AE.60 & equilibrium                          \\
                          &      &     &       & 10 $\times$CO, 0.1$\times$CH$_{4}$   \\
                          &      &     &       & 100$\times$CO, 0.01$\times$CH$_{4}$  \\
                          & 900  & 3.8 & AE.60 & equilibrium                          \\
                          &      &     &       & 10 $\times$CO, 0.1$\times$CH$_{4}$   \\
                          &      &     &       & 100$\times$CO, 0.01$\times$CH$_{4}$  \\
\hline
\ref{cd mixed}            & 1400 & 4.0 & AE.60 & equilibrium                          \\
                          & 700  &     & A.100 &                                      \\
\hline
\ref{e normal}            & 1000 & 4.0 & AE.60 & equilibrium                          \\
\hline
\ref{e non-eq}            & 1000 & 4.0 & AE.60 & equilibrium                          \\
                          &      &     &       & 10 $\times$CO, 0.1$\times$CH$_{4}$   \\
                          &      &     &       & 100$\times$CO, 0.01$\times$CH$_{4}$  \\
\hline
\ref{e mixed}             & 1400 & 4.0 & AE.60 & equilibrium                          \\
                          & 700  &     & A.100 &                                      \\
\enddata
\tablecomments{Cloud types (`A' and `AE) refer to cloud thickness and the associated numbers (`60' and `100') refer to modal dust grain size, as parameterized in \citet{2011ApJ...737...34M}. Multiples of CO and CH$_{4}$ are with respect to equilibrium chemistry models.  All models assume solar metallicity.  Further model details are discussed in \citet{2006ApJ...640.1063B} and \citet{2011ApJ...737...34M}.
}
\label{model names}
\end{deluxetable}

\clearpage

\clearpage

\clearpage

\clearpage

\clearpage

\clearpage

\clearpage

\clearpage

\clearpage

\clearpage

\clearpage

\clearpage

\clearpage

\bibliographystyle{apj}
\bibliography{database}

\end{document}